\RequirePackage[T1]{fontenc}
\RequirePackage[utf8]{inputenc}

\documentclass[final,5p,times]{elsarticle}
\pdfoutput=1

\usepackage[subpreambles=false]{standalone}
\usepackage{import}
\usepackage{array,multirow,graphicx}
\usepackage{paralist}
\usepackage{color,soul}
\usepackage[font=small,labelfont=bf,labelsep=period]{caption}
\usepackage{dblfloatfix}
\usepackage{mathtools}
\usepackage{amssymb}
\usepackage{siunitx}
\usepackage{accents}
\usepackage{bm}
\usepackage{lipsum}
\usepackage{enumitem}
\usepackage[utf8]{inputenc}
\usepackage{float}
\usepackage{amsmath}
\usepackage{dcolumn}
\usepackage{xcolor}
\usepackage{xfrac}
\usepackage[colorlinks=true, allcolors=blue]{hyperref}
\usepackage{chemformula}
\usepackage{siunitx}
\usepackage{blindtext}
\usepackage{url}

\usepackage{booktabs}
\usepackage{makecell}
\renewcommand\theadfont

\DeclareSIUnit\oersted{Oe}
\DeclareSIUnit\emu{emu}
\DeclareSIUnit\molatom{mol_{atom}}
\DeclareSIUnit\bohr{\mu_B}
\DeclareSIUnit\atom{atom}
\def\muSR{$\mu$SR}
\setlist[itemize]{align=left}

\biboptions{numbers,sort&compress}

\graphicspath{{figures/}}

\begin{document}
\begin{frontmatter}
\title{Tuning the Magnetic Properties of the CrMnFeCoNi Cantor Alloy}
\author[1]{Timothy A.\ Elmslie}
\author[2]{Jacob Startt}
\author[3]{Yang Yang}
\author[3]{Sujeily Soto-Medina}
\author[5]{Emma Zappala}
\author[1,4]{\\Mark~W.\ Meisel}
\author[3]{Michele V.\ Manuel}
\author[5]{Benjamin A. Frandsen}
\author[2]{R\'emi Dingreville}
\author[1]{James J.\ Hamlin}
\ead{jhamlin@ufl.edu}

\address[1]{Department of Physics, University of Florida, Gainesville, FL 32611-8440, USA}
\address[2]{Center for Integrated Nanotechnologies, Sandia National Laboratories, Albuquerque, NM, 87185, USA}
\address[3]{Department of Materials Science and Engineering, University of Florida, Gainesville, FL 32611-6400, USA}
\address[5]{Department of Physics and Astronomy, Brigham Young University, Provo, UT 84602, USA}
\address[4]{National High Magnetic Field Laboratory, University of Florida, Gainesville, Florida 32611-8440, USA}

\date{\today}

\begin{abstract}
    Magnetic properties of more than twenty Cantor alloy samples of varying composition were investigated over a temperature range of 5~K to 300~K and in fields of up to 70~kOe using magnetometry and muon spin relaxation.
    Two transitions are identified:  a spin-glass-like transition that appears between \SI{55}{\kelvin} and \SI{190}{\kelvin} depending on composition, and a ferrimagnetic transition that occurs at approximately \SI{43}{\kelvin} in multiple samples with widely varying compositions.
    The magnetic signatures at \SI{43}{\kelvin} are remarkably insensitive to chemical composition.
    A modified Curie-Weiss model was used to fit the susceptibility data and to extract the net effective magnetic moment for each sample.
    The resulting values for the net effective moment were either diminished with increasing \ch{Cr} or \ch{Mn} concentrations or enhanced with decreasing \ch{Fe}, \ch{Co}, or \ch{Ni} concentrations.
    Beyond a sufficiently large effective moment, the magnetic ground state transitions from ferrimagnetism to ferromagnetism.
    The effective magnetic moments, together with the corresponding compositions, are used in a global linear regression analysis to extract element-specific effective magnetic moments, which are compared to the values obtained by ab-initio based density functional theory (DFT) calculations.
    These moments provide the information necessary to controllably tune the magnetic properties of Cantor alloy variants.
\end{abstract}

\end{frontmatter}

\section{Introduction}
Work on high-entropy alloys began in the early 2000s with Cantor, Kim and Warren~\cite{cantor_novel_2002}.
However, the name ``Cantor alloy'' did not come about until a 2004 study by  Yeh~\emph{et al.}~\cite{yeh_nanostructured_2004} coined the term.
Since that time, high-entropy alloys have garnered increasing attention for opening  a massive number of yet-unexplored alloy systems for investigation.
Already, studies have found a number of high-entropy alloys with interesting and potentially useful properties such as high hardness and resistance to anneal softening~\cite{yeh_nanostructured_2004}, shape memory effects~\cite{firstov_high_2015,chen_shape_2019}, and superconductivity~\cite{motla_boron_2022}.

Much work has been done regarding the mechanical properties of Cantor alloys~\cite{cantor_microstructural_2004,otto_influences_2013,laurent-brocq_insights_2015,jang_high-temperature_2018,kim_mechanical_2018,kim_-situ_2018}, as well as the magnetic properties of similar compounds~\cite{wang_novel_2007,zhang_annealing_2010,kao_electrical_2011,lucas_magnetic_2011,singh_effect_2011,liu_microstructure_2012,ma_effect_2012,zhang_effects_2012,tsai_physical_2013,zhang_high-entropy_2013,leong_effect_2017,zuo_tailoring_2017}, and previous studies on the equiatomic Cantor alloy specifically have revealed the presence of two magnetic transitions.
The higher temperature transition occurs at approximately \SI{85}{\kelvin}, and has been identified as spin-glass-like through \muSR\ studies~\cite{elmslie_magnetic_2022}.
At approximately \SI{43}{\kelvin}, the material transitions again, this time into a ferrimagnetic state, as determined by density functional theory (DFT) simulations and experimental analysis of the magnetic entropy~\cite{elmslie_magnetic_2022}.
Despite the small size of the ferrimagnetic transition, prior work has suggested that it is not due to an impurity phase~\cite{elmslie_magnetic_2022}.
Other studies have observed notable differences in the size and temperature of these transitions, likely due to the significant processing dependence of the Cantor alloy~\cite{jin_tailoring_2016,schneeweiss_magnetic_2017,kamarad_effect_2019}.
Not only are the magnetic properties of the alloy highly sensitive to annealing and cold-working~\cite{schuh_mechanical_2015,otto_decomposition_2016,li_accelerated_2018}, high-temperature anneals do not eliminate the effects of cold-working.
Analysis which combined modified Curie-Weiss fitting, specific heat measurements, DFT calculations and Hall measurements has also revealed a large Stoner enhancement factor~\cite{elmslie_magnetic_2022}.

Additionally, few papers have reported the magnetic properties of four-element combinations of the Cantor alloy, but the nature of the magnetic states has not been universally established. For example, Kao \emph{et al.}~\cite{kao_electrical_2011} and 
Lucas \emph{et al.}~\cite{lucas_magnetic_2011} both investigated \ch{CrFeCoNi}, but Lucas \emph{et al.} state that the compound is paramagnetic at low temperature, while Kao \emph{et al.} claim that it is ferromagnetic.
In this work, a number of four-element compounds were synthesized and measured alongside numerous five-element compounds with varying compositions in order to resolve these issues.

Our measurements of magnetic susceptibility on more than twenty Cantor alloy samples with varying compositions reveal that increasing the relative amount of ferromagnetic elements, such as \ch{Fe}, enhances the overall magnetization, while larger quantities of antiferromagnetic elements decrease the magnetization.
This observation is consistent with previous Cantor alloy investigations, which reported \ch{Fe}, \ch{Co}, and \ch{Ni} tend to align ferromagnetically  while \ch{Cr} and \ch{Mn} are typically antiferromagnetic~\cite{zuo_tailoring_2017,schneeweiss_magnetic_2017,elmslie_magnetic_2022}.
Herein, these trends are quantified by least squares analysis performed using effective moments of each of the measured samples, revealing that element-specific magnetic moments can be estimated for samples that are close to the equiatomic composition.
However, the magnetic properties change when the effective moment becomes sufficiently large due to a lack of antiferromagnetic elements or a relative abundance of ferromagnetic ones.
At this point, the compound becomes ferromagnetic, rather than ferrimagnetic at low temperature as evidenced by muon spin-relaxation studies on select samples in the Mn series.
Our experimental results, combined with ab-initio based density functional theory (DFT) calculations, provide a complete picture of the compositional dependence of the magnetic properties of the Cantor alloy system.
The work highlights the great promise of similar 
regression analysis methods applied to several chemically substituted alloy samples to provide insights into elemental specific contributions to properties in other high-entropy materials.

\section{Methods}
This section describes the details of the various experimental and computational methods that were employed in this work, including the sample synthesis, magnetic and structural characterization, and density functional theory calculations.

\subsection{Synthesis}
Samples were synthesized by combining stoichiometric \linebreak 
quantities of elemental Cr, Mn, Fe, Co, and Ni in an Edmund B\"uhler MAM-1 Compact Arc Melter.
Chromium, manganese, iron, and nickel were sourced from Alfa Aesar, while cobalt was purchased from Cerac.
The chromium used for synthesis was 99.995\% pure, while the cobalt was 99.5\% pure.
All other elements were 99.95\% pure.
Each sample was melted five times, flipping it over between each melt to improve sample homogeneity.
Following synthesis, samples were sealed in quartz tubes in \ch{Ar} atmosphere and annealed at \SI{1100}{\celsius} for six days, then quenched in water before measurement.
However, the samples consisting of 25\% of one element and 18.75\% of each other element (25\% samples) were annealed first for six days at \SI{1080}{\celsius} and then again at \SI{1100}{\celsius}, quenching in water following each anneal.

\subsection{Sample Nomenclature}
To simplify the process of investigating varying compositions, in each sample, the amount of only one element was adjusted, and all other elements remained in an equiatomic ratio according to the atomic ratios, A$_x$[BCDE]$_{100-x}$.
For example, when adjusting the amount of \ch{Mn}, sample compositions would follow the pattern \ch{Mn}$_x$(\ch{CrFeCoNi})$_{100-x}$.
This practice permits a simplified notation, listing only the element with an adjusted quantity and the percentage of that element present in the alloy.
From that information alone, the relative proportion of other elements can then be inferred.
For example, ``Cr$_{22}$'' is used instead of writing ``\ch{Cr22(MnFeCoNi)78}'', and a sample with composition \ch{CrFeCoNi} is represented by the abbreviation ``Mn$_0$.''
This convention is used from this point onward in this work.
Samples with an equal proportion of each element (equiatomic) are listed as ``Equi.''

\subsection{Magnetic Characterizations}
Magnetization and susceptibility measurements were performed using a Quantum Design Magnetic Property Measurement System (MPMS).
Small pieces ranging from a few milligrams to a few hundred milligrams (typically $\approx$ \SI{50}{\milli \gram}), were cut from larger samples using an Allied 3000 low speed saw to minimize unintentional working of the samples.
Each sample was secured in a gel capsule inside a plastic straw for measurement.

Muon spin spectroscopy measurements (\muSR) were conducted at \textsc{TRIUMF} Laboratory in Vancouver, Canada using the \textsc{LAMPF} spectrometer on the M20D beamline. 
The \muSR\ technique utilizes the asymmetric decay of initially spin-polarized positive muons into positrons to probe the magnetic properties of the sample~\cite{hilli;nrmp22}. 
Specifically, \muSR\ is uniquely sensitive to the volume fraction of different magnetic and nonmagnetic phases, which is important for potentially inhomogeneous systems such as magnetic alloys. 
The experimentally measured quantity is the time-dependent asymmetry, $a(t)$, which is determined from the difference in positron events between two detectors placed on opposite sides of the sample. 
The asymmetry is proportional to the component of the net muon spin polarization along the axis connecting the two detectors. 
Information about the local magnetic field distribution can be inferred from the behavior of $a(t)$ as a function of temperature and applied magnetic field. 
Additional details about our experimental configuration are described elsewhere~\cite{elmslie_magnetic_2022}, and 
the open source program BEAMS~\cite{peter;gh21} was used for data analysis. 

\subsection{Structural and Composition Characterizations}
The crystal structure of the sample was investigated using a Panalytical Xpert X-Ray Diffraction (XRD) with a Cu K-alpha radiation source at a voltage of \SI{45}{\kilo\volt} and current of \SI{40}{\milli\ampere} in the 2$\theta$ range of 30-\ang{100}. 
The powder samples were prepared using a steel file. 
The microstructure of the sample was characterized using a Tescan MIR3 scanning electron microscope (SEM) at \SI{20}{\kilo\volt}. 
The expoxy-mounted sample was first ground using a series of SiC paper at steps of 600 grit, 800 grit, and 1200~grit, then further polished using alcohol-based lubricant and diamond paste in the sequence of \SI{6}{\micro\meter}, \SI{3}{\micro\meter}, and \SI{1}{\micro\meter}. 
The sample surface was master polished using \SI{0.05}{\micro\meter} water-free colloidal silica suspension. 
The chemical composition of the sample was determined using an equipped energy-dispersive X\nobreak{-}ray (EDX) detector. 
\subsection{Computational methods}
Ab-initio based density functional theory (DFT) calculations were performed in parallel to experimental investigation to study the net moment behavior of the low temperature ferrimagnetic phase as a function of composition. 
All DFT calculations were performed in the spin-collinear polarized state using the Vienna Ab-initio Simulation Package (VASP)~\cite{Kresse1993, Kresse1994, Kresse1995}.  
Electronic wavefunctions were thus modeled using plane-waves while the interaction between the frozen core states and valence states was handled according to the projector-augmented wave (PAW) formalism~\cite{Blochl1994, Kresse1999}. 
Exchange and correlation was treated according to the generalized gradient approximation as parameterized by Perdew, Burke, and Ernzerhof (PBE)~\cite{Perdew1996}. 

The modeled Cantor compositions were selected according to the same convention used in the experimental analysis, such that only one element was varied at time, according to A$_x$[BCDE]$_{100-x}$, resulting in five separate series of simulations (\emph{i.e.,} one for each species). 
Within each series, four compositions were modeled (\emph{i.e.,} $x=10$, 20, 30, 40). 
The alloys were modeled using supercells containing 108 atoms, constructed as $3\times3\times3$ cubic multiplications of the conventional 4~atom FCC unit cell. 
Three separate supercells, each with a newly randomized ordering of the atoms among the lattice sites, were constructed for each composition modeled. 
Thus, any reported properties represent the average of the three random solution supercells at each composition. 

For all calculations, a $5\times5\times5$ gamma-centered k-point mesh, along with a plane-wave energy cutoff of \SI{400}{\eV} and a Gaussian smearing smearing width of \SI{0.01}{\eV} was found to sufficiently minimize fluctuations in the total energy and magnetic moments. The minimization threshold of the total energy with respect to the electronic convergence was set to \SI{1.0E-6}{\eV}, while ionic convergence threshold was met when forces on all atoms fell below \SI{20E-3}{\eV \per \angstrom}.

\section{Results}
In this section, the results of magnetization as a function of temperature and magnetic field are described along with the low temperature \muSR\ data. 
These experimental findings are analyzed and contrasted with the DFT work in the next section. 

\subsection{Magnetization}
\begin{figure*}[ht!]
\centering
\includegraphics[width=14cm]{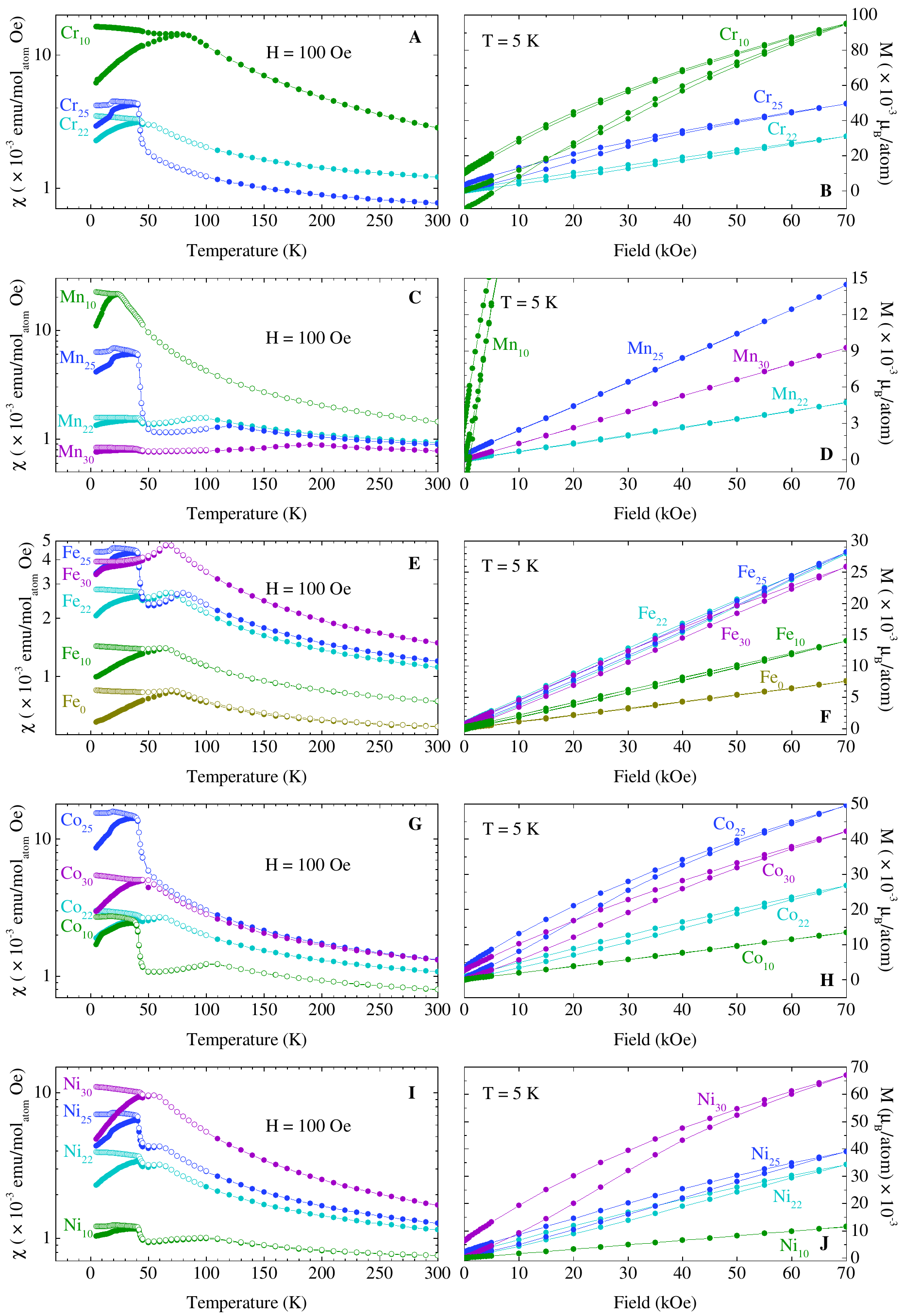}
    \caption{Susceptibility and magnetization data for Cantor alloys of varying compositions. 
    Panels on the left show susceptibility versus temperature on a logarithmic scale, while those on the right show magnetization as a function of magnetic field.  
    In the susceptibility plots, solid circles represent zero-field cooled (ZFC) data, which were measured after cooling down in zero field.
    Open circles represent field-cooled warming (FCW) data, which were collected after cooling down in an applied field of \SI{70}{\kilo \oersted}.  
    Each row displays a different set of compounds: (A)-(B) Cr-varied compositions, (C)-(D) Mn-varied compositions, (E)-(F) Fe-varied compostions, (G)-(H) Co-varied compositions, and (I)-(J) Ni-varied compositions.  Section 2.1.2 describes the sample nomenclature.  Certain curves such as Mn$_{10}$ and Fe$_{30}$ display only one feature.  In these cases, the feature is identified as $T_1$ or $T_2$ depending on its shape, with $T_1$ being step-like and $T_2$ being a peak.}
    \label{fig:Sub_All}
\end{figure*}
\begin{figure}
    \centering
    \includegraphics[width=\columnwidth]{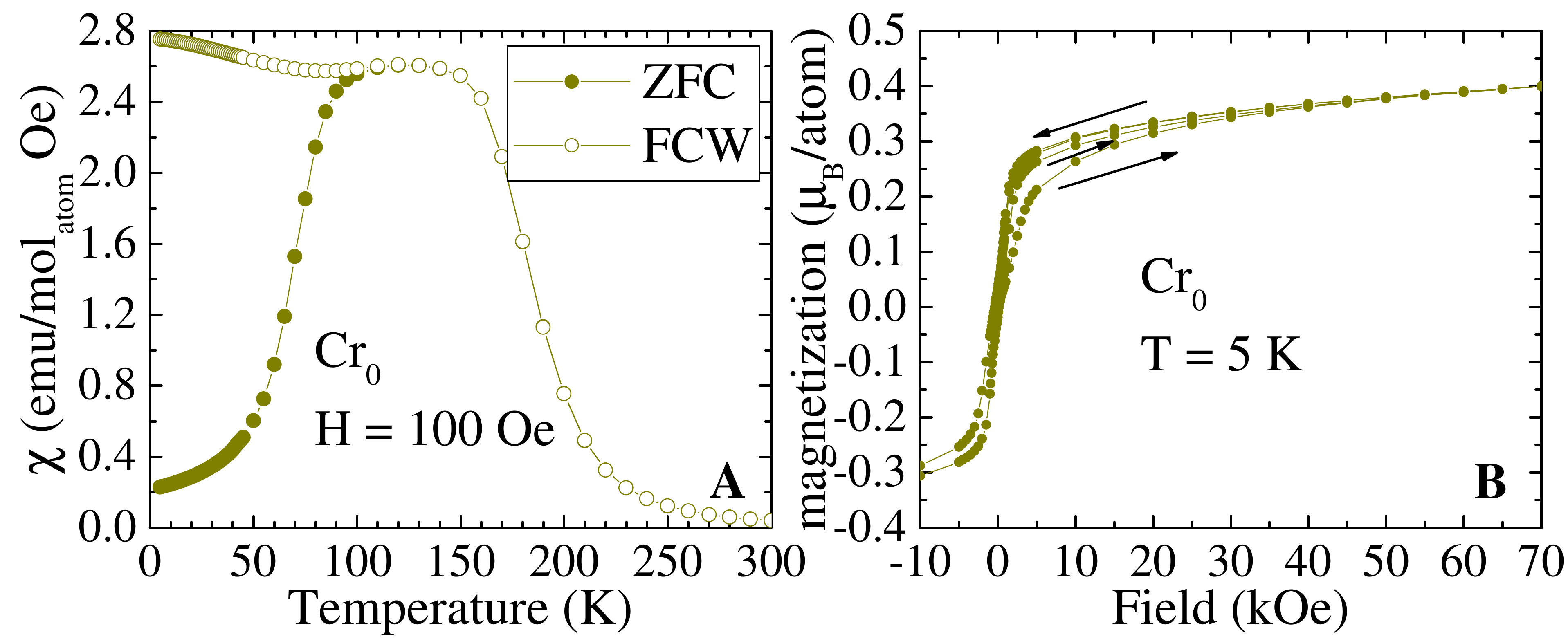}
    \includegraphics[width=\columnwidth]{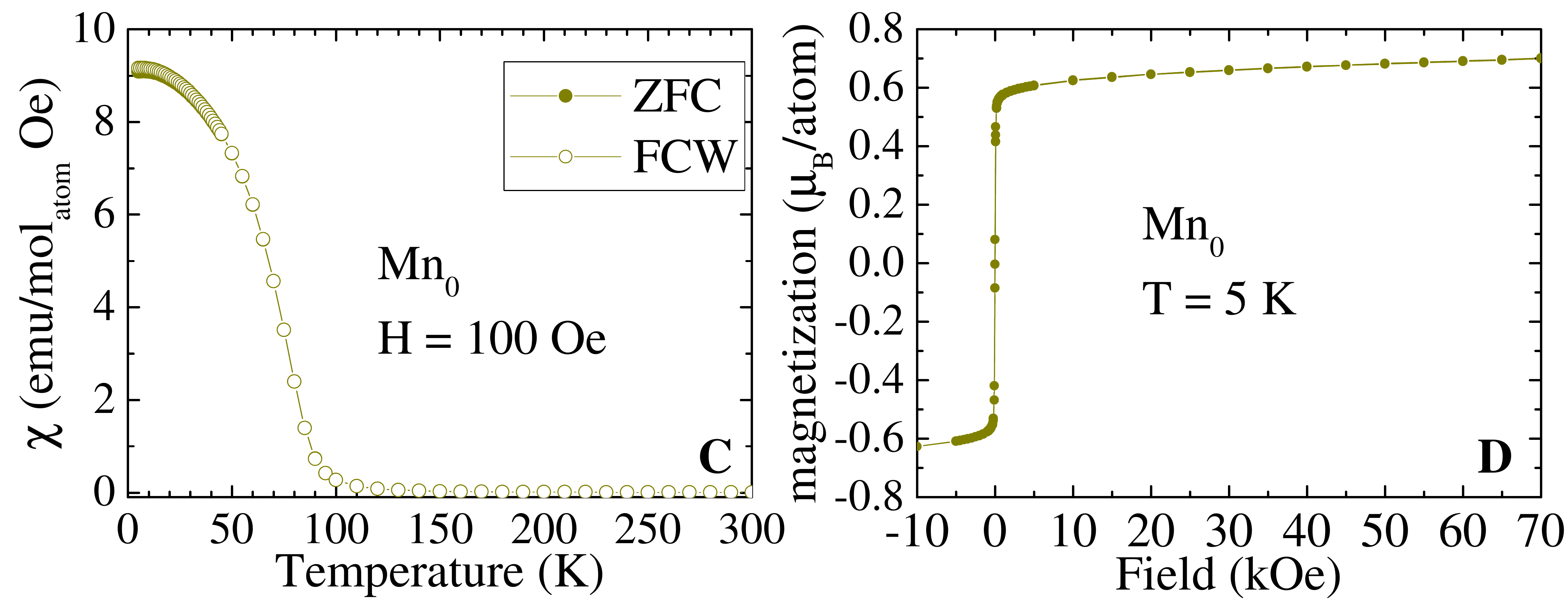}
    \caption{Magnetic susceptibility as a function of temperature for the (A) Cr$_0$ and (C) Mn$_0$ samples. Magnetization as a function of applied field for the (B) Cr$_0$ and (D) Mn$_0$ samples.  
    Magnetization data for Cr$_0$ show significant hysteresis, and arrows in panel (B) indicate the direction of the magnetic field sweep for each curve segment.  
    In panel (C), ZFC and FCW curves are difficult to distinguish due to significant overlap, with less than 1\% of difference across the entire data range.}
    \label{fig:Cr0}
\end{figure}

In Fig.~\ref{fig:Sub_All}, results of susceptibility versus temperature measurements are presented on the left side, while magnetization versus field results are shown on the right.
Each row displays a collection of samples, grouped according to the elemental concentration being adjusted.
Due to their strong magnetic response, most of the $x = 0$ samples are not included in this figure, with Fe$_0$ as an exception, so the composition dependence of the trends and features of the $x > 0$ data can be more easily compared.  
The Cr$_0$ and Mn$_0$ data sets are plotted in Fig.~\ref{fig:Cr0}.

Three striking features are identified in the plots shown in Fig.~\ref{fig:Sub_All}.
Firstly, a spin-glass-like transition ($T_2$), marked by a peak~\cite{schneeweiss_magnetic_2017,elmslie_magnetic_2022}, occurs at high temperatures, which varies widely depending on composition.
This feature is most apparent in the Fe$_{30}$ dataset at about \SI{67}{\kelvin}, Fig.~\ref{fig:Sub_All}E.
Secondly, a step-like feature ($T_1$) appears in many datasets, marking a ferrimagnetic transition~\cite{elmslie_magnetic_2022}, as clearly present in the Mn$_{25}$ susceptibility data, Fig.~\ref{fig:Sub_All}C.
In all samples in which it appears, the $T_1$ feature occurs at \SI{43(1)}{\kelvin}, and although this robustness might be considered as a fingerprint of an impurity phase, prior work has established this transition to be intrinsic to the Cantor alloy~\cite{elmslie_magnetic_2022}.
Lastly, another feature ($T^*$) occurs in several samples at approximately \SI{18}{\kelvin} and is most obvious in the Mn$_{25}$ data, Fig.~\ref{fig:Sub_All}C. 
The nature of the $T^*$ feature is unclear, but may be the result of differences in processing.
The $x=25$ samples were annealed and quenched twice, once at \SI{1080}{\kelvin} and then at \SI{1100}{\kelvin}, whereas other samples in which $T^*$ appears were annealed and subsequently quenched in the same quartz tube along with several more pieces (up to as many as five), which may have resulted in a slower quench.

The magnetization versus field data along the right side of Fig.~\ref{fig:Sub_All} shows that even at fields of up to \SI{70}{\kilo \oersted}, magnetization remains a small fraction of a Bohr magneton.
For example, the Fe$_{25}$ sample reaches approximately \SI{0.03}{\bohr \per \atom}, much smaller than the Hund's-rule-predicted effective moments for the constituent elements, which is typically a few \si{\bohr \per \atom}~\cite{blundell_magnetism_2001}.
Thus, near-equiatomic compositions of the Cantor alloy are far from magnetic saturation saturation even at 70~kOe.
Even in Mn$_{10}$, which extends beyond the limits of the y-axis (Fig.~\ref{fig:Sub_All}D), the magnetization reaches only \SI{0.08}{\bohr \per \atom} at \SI{70}{\kilo \oersted}.
One trend visible in Fig.~\ref{fig:Sub_All} is that hysteresis increases as susceptibility increases, perhaps indicating the presence of ferromagnetic regions within the material.

The plots of Fig.~\ref{fig:Sub_All} also reveal that susceptibility increases with greater proportions of ferromagnetic elements \ch{Fe}, \ch{Co}, or \ch{Ni}.
On the other hand, increasing the relative amount of antiferromagnetic elements \ch{Cr} or \ch{Mn} decreases the susceptibility.
At sufficiently low concentrations of \ch{Cr} or \ch{Mn}, the magnetic properties change significantly, resulting in larger saturation magnetization and susceptibility.
Consequently, \ch{Cr0} and \ch{Mn0} data are excluded from Fig.~\ref{fig:Sub_All} and are instead presented in Fig.~\ref{fig:Cr0}.
The Cr$_0$ sample reaches susceptibility values of approximately \SI{9.1}{\emu \per \molatom \per \oersted}, much larger than the susceptibility maximum of, for example, Mn$_{25}$ of about $6.8 \times 10^{-3}$ emu mol$_{\rm atom}^{-1}$~Oe$^{-1}$.
Although the susceptibility and saturation magnetization of the Cr$_0$ and Mn$_0$ samples are orders of magnitude larger than those shown in Fig.~\ref{fig:Sub_All}, they are still lower than effective moments predicted by Hund's rule~\cite{blundell_magnetism_2001}.
These Cr$_0$ and Mn$_0$ samples also possess a single ferromagnetic transition, compared to the ferrimagnetic and spin-glass-like transition of the near-equiatomic samples.

Transition temperatures for each measured sample are summarized in Table~\ref{tab:sample_table}.
The positions of the lower temperature (``step-like'') transitions were determined by local minima of the first derivative, while the critical temperature of the higher temperature transitions are located at the peak in the magnetic susceptibility.
In several of the iron samples, the ferrimagnetic transition is either absent are too small to detect, and the spin-glass/antiferromagnetic transition is not visible in the ferromagnetic (Cr$_0$ and Mn$_0$) or nearly ferromagnetic (Mn$_{10}$) samples.
\begin{table*}[htb]
    \centering
    \small
    \sisetup{table-format = 3.2}
    \begin{tabular}{
    l
    r
    r
    S[table-number-alignment = center]
    S[table-number-alignment = center]
    S[table-number-alignment = center]
}
Composition
& {\quad\quad $T_1$ (K)}
& {\quad\quad $T_2$ (K)}
& {\quad\quad\quad $\chi_0 (\times 10^{-3} \mathrm{emu}\, \mathrm{mol}_{\mathrm{atom}}^{-1} \mathrm{Oe}^{-1})$}
& {\quad\quad\quad $p_{\rm eff}$ ($\mu _B$)}
& {\quad\quad\quad $\theta$ (K) \quad\quad} \\
\toprule
Equi ($^{\S}$) & 44 & 82 & 5.57(4) & 0.98(10) & -8.9(8) \\
\midrule
Cr$_{0}$ ($^{* \dagger}$) & 180 & {-} & {-} & {-} & {-} \\
Cr$_{10}$ ($^*$) & 44 & 80 & -10.1(5) & 2.9(4) & 15.6(9) \\
Cr$_{22}$ & 44 & 55 & 7.761(16) & 1.02(6) & -5.10(21) \\
Cr$_{25}$ & 42 & {-} & 5.287(22) & 0.77(7) & -9.6(5) \\
\midrule
Mn$_{0}$ ($^{* \dagger})$ & 75\rlap{$^{\ddag}$} & {-} & {-} & {-} & {-} \\
Mn$_{10}$ & 44 & {-} & 1.75(23) & 1.67(19) & 13.3(5) \\
Mn$_{22}$ & 43 & 100 & 5.640(27) & 0.95(9) & -11.6(9) \\
Mn$_{25}$ & 42 & 120 & 5.608(28) & 0.91(10) & -17(1) \\
Mn$_{30}$ & 43 & 190\rlap{$^{\ddag}$} & 6.07(27) & 0.6(3) & 14(29) \\
\midrule
Fe$_{0}$ & {-} & 70 & 4.667(23) & 0.44(7) & 10(2) \\
Fe$_{10}$ & {-} & 65 & 5.281(20) & 0.73(7) & -11.7(8) \\
Fe$_{22}$ & 43 & 67 & 5.868(9) & 1.13(5) & -5.20(14) \\
Fe$_{25}$ & 42 & 80 & 5.80(3) & 1.23(10) & -7.2(5) \\
Fe$_{30}$ & {-} & 67 & 5.09(14) & 1.51(18) & 3(1) \\
\midrule
Co$_{10}$ & 42 & 105 & 5.264(16) & 0.83(7) & -13.6(7) \\
Co$_{22}$ & 43 & 62 & 6.079(13) & 1.07(6) & -4.37(19) \\
Co$_{25}$ & 42 & {-} & 4.93(5) & 1.39(10) & 3.4(3) \\
Co$_{30}$ & {-} & 50 & 5.60(3) & 1.34(8) & 0.15(21) \\
\midrule
Ni$_{10}$ & 43 & 100 & 6.35(3) & 0.53(10) & 9(3) \\
Ni$_{22}$ & 43 & 60 & 5.653(17) & 1.18(6) & -4.20(19) \\
Ni$_{25}$ & 42 & 60 & 4.44(3) & 1.40(9) & -1.27(24) \\
Ni$_{30}$ & 44 & 55 & 0.05(17) & 1.96(19) & 10.0(6) \\
\bottomrule
\end{tabular}

    \caption{Transition temperatures and modified Curie-Weiss fitting parameters of each measured alloy.
    $^{\S}$Data on the equiatomic compound were obtained by averaging values from four samples.
    $^{*}$Asterisks mark samples that could not be accurately fit by the modified Curie-Weiss equation or produced unphysical results from fitting.  
    $^{\dagger}$Daggers indicate samples which become ferromagnetic below the $T_1$ transition temperature, as opposed to the ferrimagnetic state assumed by other samples.
    $^{\ddag}$Double daggers indicate transition temperatures that were confirmed via muon spin-relaxation measurements.}
    \label{tab:sample_table}
\end{table*}

\subsection{Muon Spin Spectroscopy}
\begin{figure}
    \centering
    \includegraphics[width=0.45\textwidth]{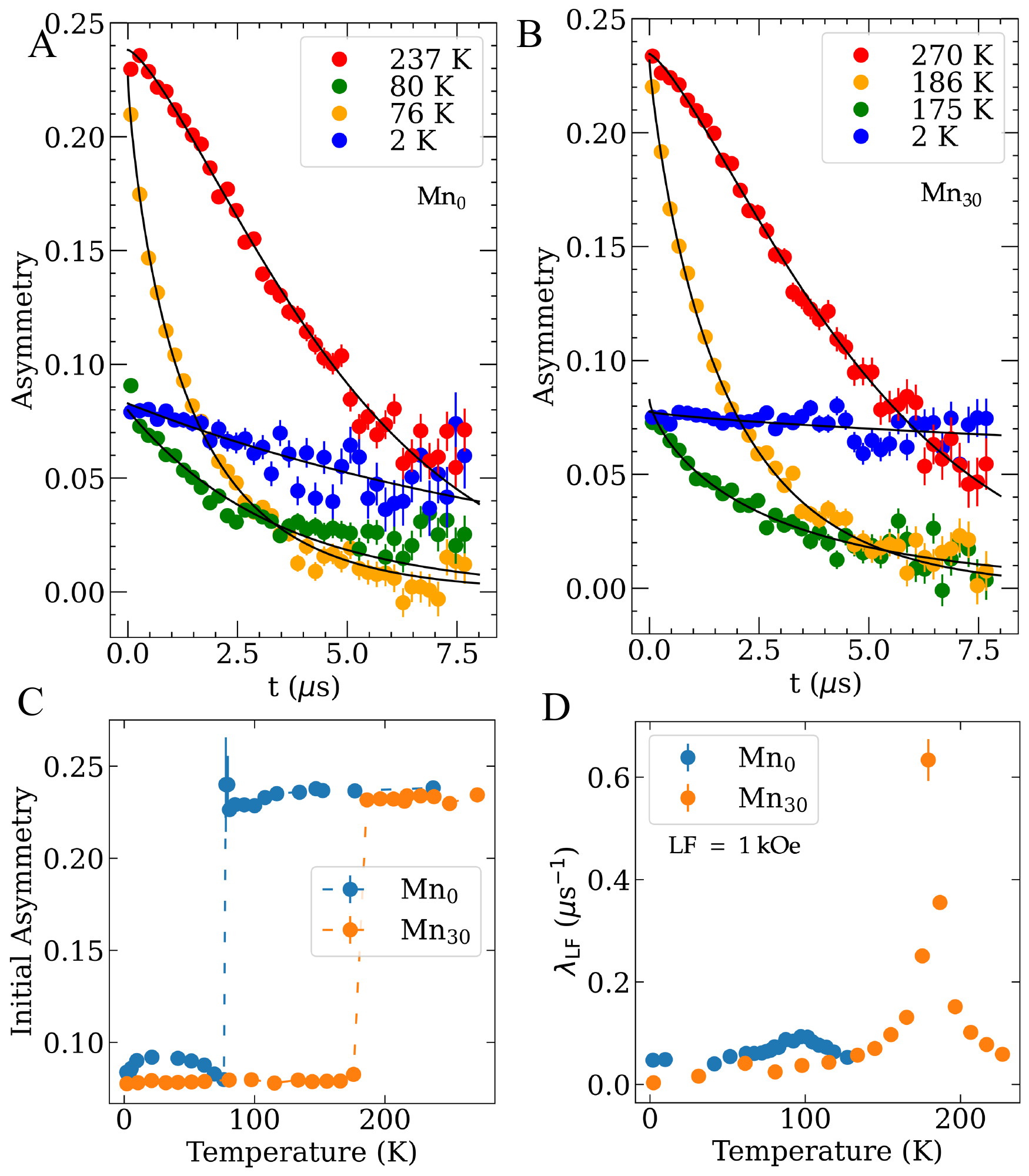}
    \caption{(A, B) Representative zero-field \muSR\ asymmetry spectra for Mn$_0$ and Mn$_{30}$, respectively. 
    (C) Temperature dependence of the initial asymmetry at $t=0$ for both samples, showing a sharp step at the expected transition temperatures. 
    (D) Temperature dependence of the long-time relaxation rate $\lambda_{\mathrm{LF}}$ in an applied longitudinal field of 1~kOe.}
    \label{fig:muSR}
\end{figure}

\muSR\ measurements of Mn$_0$ and Mn$_{30}$ alloys were conducted to evaluate the magnetic homogeneity and confirm the magnetization results. 
Representative $a(t)$ asymmetry spectra collected in zero applied magnetic field (ZF) are shown in Fig.~\ref{fig:muSR}(a) and (b) for Mn$_0$ and Mn$_{30}$, respectively.
For both samples, the gentle relaxation of the asymmetry at high temperature becomes more rapid as the temperature is lowered toward the transition due to critical slowing down of spin fluctuations. 
Between $77 - 80$~K for Mn$_0$ and $175 - 185$~K for Mn$_{30}$, the initial asymmetry value at $t=0$ drops to approximately 1/3 of  its high-temperature value, signifying the onset of static magnetism~\cite{hilli;nrmp22}. 
These temperatures agree well with $T_1 = 75$~K for Mn$_0$ and $T_2 = 190$~K for Mn$_{30}$ given in Table~\ref{tab:sample_table}.
The narrow temperature range in which the initial asymmetry drops demonstrates that the magnetic transition is sharp and uniform throughout the entire volume of the sample, \emph{confirming that the non-stoichiometric alloys are similarly magnetically homogeneous as the equiatomic alloy}~\cite{elmslie_magnetic_2022}.
The uniform magnetic behavior of each sample is further illustrated in Fig.~\ref{fig:muSR}(c), where the initial asymmetry plotted as a function of temperature shows an abrupt step at the transition. The initial asymmetry was determined by fitting the generalized exponential function $a(t) = a_0 \exp(-\lambda t)^{\beta}$ to the spectra, as was done previously~\cite{elmslie_magnetic_2022}.  
It is noteworthy that a characteristic Kubo-Toyabe relaxation pattern~\cite{uemur;ms99} is observed for Mn$_0$ within the first $\sim$100~ns of the spectra below the transition, but is not visible on the time scale used in Fig.~\ref{fig:muSR}(a), in contrast to Mn$_{30}$ and the equiatomic alloy~\cite{elmslie_magnetic_2022}, indicative of subtle composition-dependent changes of the internal magnetic field distribution at the muon stopping sites. 

Finally, Fig.~\ref{fig:muSR}(d) displays the temperature-dependent relaxation rate $\lambda_{\mathrm{LF}}$ of the long-time asymmetry for data collected in an applied longitudinal field (LF) of 1~kOe (not shown), which is a sensitive probe of spin dynamics. 
The values of $\lambda_{\mathrm{LF}}$ were determined from generalized exponential fits to the LF asymmetry spectra. Mn$_0$ shows a low, broad peak in $\lambda_{\mathrm{LF}}$ centered around 90~K (slightly above the transition temperature in ZF), while Mn$_{30}$ shows a much higher and sharper peak centered around 180~K. 
These results point to differences in the spin dynamics of the two samples, with critical behavior being suppressed in Mn$_0$ and preserved in Mn$_{30}$. A more detailed theoretical investigation of the microscopic magnetic behavior of Cantor alloys with various Mn-compositions could help explain these differences in the ZF and LF data of the Mn$_0$ and Mn$_{30}$ samples.

\section{Discussion}
\subsection{Modified Curie-Weiss Fitting}
Previous work~\cite{elmslie_magnetic_2022} on fitting magnetic susceptibility data for the equiatomic Cantor alloy has revealed the need for a modified Curie-Weiss equation
\begin{equation}
    \chi(T) = \chi_{\rm{o}} + \frac{C_{m}}{T-\theta}\;,
    \label{eq:modCW}
\end{equation}
in which $C_m$ is the Curie constant, $\theta$ is the Curie-Weiss 
temperature, and $\chi_{\rm{o}}$ is a constant that represents the summation of multiple temperature-independent terms such as Pauli paramagnetism, Van Vleck paramagnetism, Landau diamagnetism, and core diamagnetism~\cite{mulay_theory_1976,cao_charge_2000,kaldarar_thermal_2009,gron_influence_2011,markina_magnetic_2014}.
One can obtain an effective moment $p_{\rm{eff}}$ from this fitting using~\cite{mcelfresh_fundamentals_1994}
\begin{equation}
    p_{\rm{eff}} = 2.82\,C_m^{1/2}\;.
    \label{eq:peff}
\end{equation}
In the Cantor alloy, Stoner-enhanced Pauli paramagnetism dominates the $\chi_{\rm{o}}$ term, and diamagnetic contributions were predicted to be, at most, 5\% of the size of the enhanced Pauli paramagnetism~\cite{elmslie_magnetic_2022}.
Van Vleck paramagnetism is present in compounds with a total angular momentum  J = 0 ground state.
In the Cantor alloy, this state may occur in Cr$^{2+}$ and Mn$^{3+}$, though further work is required to determine whether this mechanism plays a significant role.

Fitting Eq.~\ref{eq:modCW} to the field-cooled susceptibility data produces best fit values of $\chi_0$, $p_{\rm{eff}}$, and $\theta$ for 23 different samples.
In each case, the data were fit from \SI{300}{\kelvin} down to temperatures just above the highest temperature magnetic ordering transition.
The parameters derived from fitting each measured sample are summarized in Table~\ref{tab:sample_table}.
Figure~S1 (supplementary material) compares the susceptibility with the fit to Eq.~\ref{eq:modCW} for each sample, demonstrating excellent agreement in almost all cases.
Fit parameters are not reported for the two samples that appear to have transitioned to a distinct ferromagnetic ground state, Mn$_0$ and Cr$_0$.
For these two samples the fit was unreliable, producing values that were unphysical or depended significantly on the temperature range of the fit.

Several conclusions are immediately apparent on review of Table~\ref{tab:sample_table}.
While the values of the Curie-Weiss temperature, $\theta$, remain close to zero, systematic trends are absent.
The temperature independent susceptibility, $\chi _0$, is consistently of order $ \sim \SI{5E-3}{\emu \per \molatom \per \oersted}$.
However, systematic trends in $\chi _0$ are also largely absent, except perhaps in the case of the Ni series.
On the other hand, the effective moments, $p_{\rm{eff}}$ show very clear trends across each substitutional series.
These trends are illustrated in Fig.~\ref{fig:peffs}a, where it is immediately clear that Cr and Mn suppress the susceptibility, while Fe, Co, and Ni have the opposite effect.
\begin{table}
 \centering
 \begin{center}
\begin{tabular}{p{1.5cm} l l l l l l}
 Source & $p_{\rm{Cr}}$ & $p_{\rm{Mn}}$ & $p_{\rm{Fe}}$ & $p_{\rm{Co}}$ & $p_{\rm{Ni}}$\\
 \toprule
 This work (experiment) & -5.0(1.2) & -3.1(1.2) & 3.8(3) & 3.2(4) & 6.5(7) \\
 This work (DFT) & -0.2(1) & -0.3(1) & 1.3(2) & 0.7(1) & 0.3(4) \\
 \midrule
 Ionization 2+~\cite{blundell_magnetism_2001} & 4.82 & 5.82 & 5.36 & 4.90 & 3.12 \\
 Ionization 3+~\cite{blundell_magnetism_2001} & 3.85 & 4.82 & 5.82 & - & - \\
 \bottomrule
 \end{tabular}
 \end{center}
 \caption{Effective moments and their uncertainties extracted via least-squares regression compared with empirical effective moments as tabulated in Blundell~\cite{blundell_magnetism_2001}.}
 \label{tab:peffs}
\end{table}

\subsection{Element-specific average magnetic moments}
Average element-specific magnetic moments can be estimated via a least-squares method using the effective moments in Table~\ref{tab:sample_table}.
Assuming that each element contributes to the total effective moment as a linear function of its concentration, the net effective moment of the compound can be expressed as
\begin{equation}
 p_{\rm{eff}} = \sum _i C_i\, p_i\;\;,
\end{equation}
in which $C_i$ is the concentration of element $i$, and $p_i$ is the effective moment of that element.
For example, the net effective moment of Ni$_{25}$ could be estimated as
\begin{equation}
 0.25\, p_{\rm{Ni}} + 0.1875 (p_{\rm{Cr}} + p_{\rm{Mn}} + p_{\rm{Fe}} + p_{\rm{Co}}) = p_{\rm{eff,Ni_{25}}}\;,
\end{equation}
and similar equations can be generated for each of the samples for which $p_{\rm{eff}}$ values are given in Table~\ref{tab:sample_table}.
If one assumes the element-specific magnetic moments are constant across all samples, an over-determined system of equations with only five unknowns is obtained.
This set of linear equations can be solved by linear regression to produce best-fit values of the five element-specific magnetic moments.
Additional details of the fitting process are described in the Supplementary Material, Sec.~S1.

\begin{figure}[h]
 \centering
 \includegraphics[width=0.8\columnwidth]{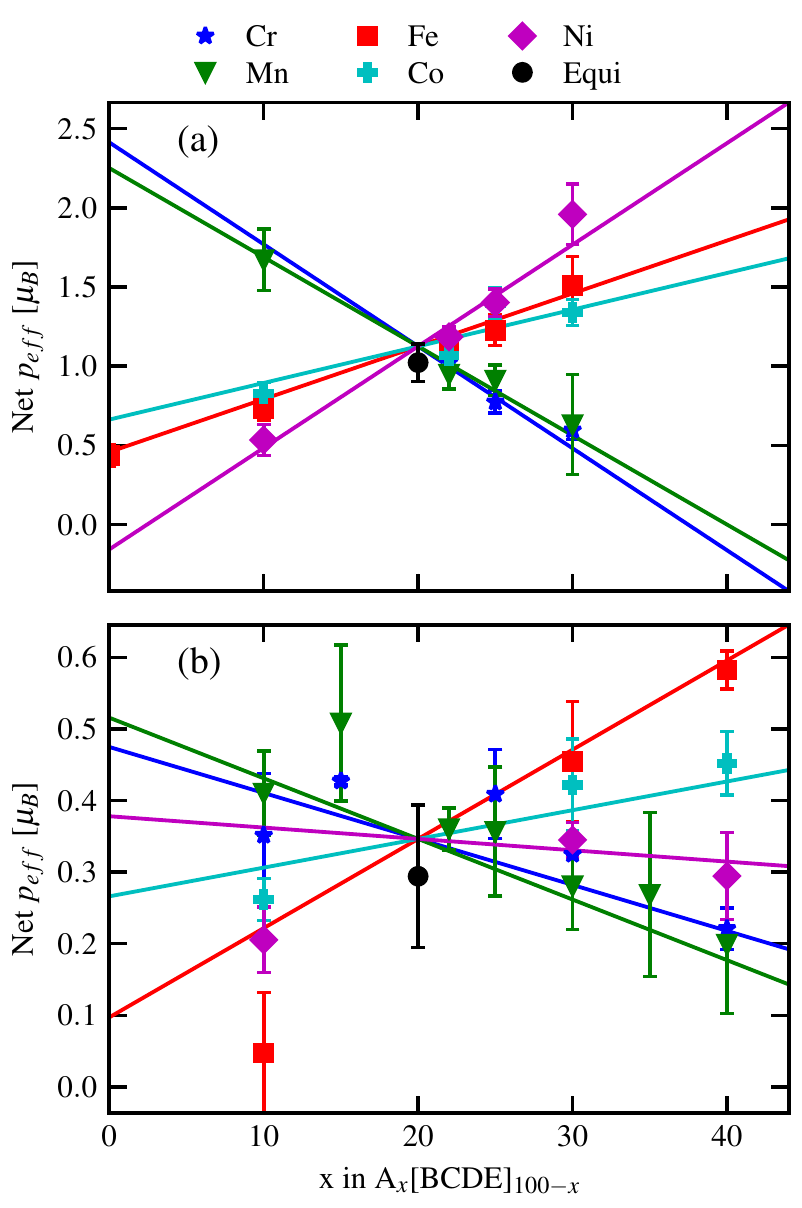}
 \caption{Net magnetic moment vs composition for each series of samples for (a) experimental data and (b) DFT simulations.  Each point corresponds to a single sample.
 Adding Fe, Co, or Ni tends to increase the net moment, while adding Cr or Mn suppresses the net moment.
 The solid lines represent a model of the data based on element-specific $p_{\rm{eff}}$ values listed at the bottom of the figure.
 The element-specific $p_{\rm{eff}}$ values were determined through a global linear least squares regression of all of the available data as described in the text.}
 \label{fig:peffs}
\end{figure}

The resulting element-specific moments are tabulated in Table~\ref{tab:peffs}.
The extracted moments are of order a few $\mu _B$, similar to typical experimental values found for these elements and also of the same order as the Hund's rules values~\cite{blundell_magnetism_2001}.
However, the Cr and Mn moments enter as negative values, suggesting these elements exhibit a strong antiferromagnetic interaction.
The solid lines in Fig.~\ref{fig:peffs} are generated for each substututional series using the moments listed in Table~\ref{tab:peffs} together with Eq.~\ref{eq:peff}.
Strikingly, this single set of five element-specific moments produces excellent agreement with the experimentally determined net effective moment across a wide range of compositions.

\subsection{Comparison of DFT and experiment}
Previous DFT calculations of the equiatomic Cantor alloy suggested a ferrimagnetic ordering at absolute zero \cite{elmslie_magnetic_2022}, where the Fe, Co, and Ni species, and the Cr and Mn species couple ferromagnetically and antiferromagnetically, respectively.
In this work, the compositional dependence of the net moment was modeled in parallel to the experimental study, probing the dependence on a per species basis in the same manner as the experimental investigation.
These results are shown in Fig.~\ref{fig:peffs}(b), where the net moment, as calculated by DFT, is plotted as a function of the variation of each species in the alloy, such that A$_x$[BCDE]$_{100-x}$ represents the overall composition.
In a similar fashion, a least-square fit is applied to the set of DFT calculated moments to obtain effective local moments of the individual species (see Table~\ref{tab:peffs}).
Both DFT and experimentally derived element specific moments exhibit consistent signs, i.e., they are in agreement on the ferro- and antiferromagnetic tendency of each species. 
However, the magnitudes of the effective moments differ.
Specifically, they differ on the ferromagnetic strengths of Fe and Ni and on the relative antiferromagnetic strength between Cr and Mn.
Notably, DFT calculations predict a weak decrease in net effective moment with increasing Ni concentration, contrary to experimental results.
This discord occurs despite the positive moment produced by the linear regression analysis of the DFT moments for Ni.
In fact, this behavior can be understood as being due to the rather small Ni moment compared to the moments of Co and Fe, such that increasing Ni concentration proportionally decreases Co and Fe content, which lowers the overall net moment.

There could be several reasons for the discrepancy in the element specific moments from the analysis of DFT \emph{versus} experimental data.
First, the experimental $p_{\rm eff}$ moments were fit to susceptibility data measured in the temperature range of the paramagnetic state, \emph{i.e.,} above any magnetic ordering transition temperature. 
The DFT $p_{\rm eff}$ moments, on the other hand, were calculated at absolute zero, squarely within the temperature range corresponding to the ferrimagnetic state. 
Attempting to model the compositional dependence of the paramagnetic state through DFT, however, is not computationally feasible and would likely be subject to errors common to the approximations needed to depict the paramagnetic state in DFT.

Independent of the overall magnetic state, another potential contributing factor for the discrepancy between computationally and experimentally derived moments may be related to thermal expansion.
The average volume associated with each atomic lattice site is a factor that has been shown have significant influence on the local magnetic moment behavior of species in this alloy~\cite{Ma_dft-volume_2015}.
As the volume of the lattice site increases, the magnitude of the local moment of each species may also change at different rates.
In the case of Mn, Ma~\emph{et al.}~\cite{Ma_dft-volume_2015} even showed the net direction of the average Mn spin tended to flip from down to up (or from antiferromagnetic to ferromagnetic) as the atomic volume increases.
While the total amount of thermal expansion in the low temperature regime investigated here ($T <300$~K) is likely small, effects on the local and net moment behavior may still be present.  

Billington \emph{et al.}~\cite{billington_bulk_2020} carried out a detailed study of the element specific magnetic properties of the Cantor alloy and certain Cantor alloy variants using a combination of magnetic Compton scattering, x-ray magnetic circular dichroism spectroscopy, and bulk magnetization measurements,
noting significant discrepancies between computational and experimental results when examining magnetic moments.
However, in compounds where computation and experiment disagreed, computational moments were much larger than experimental ones.
For instance, they measured a net moment of \SI{0.008}{\bohr} in a \textsc{SQUID} magnetometer for the equiatomic Cantor alloy, while different theoretical calculations predicted moments between 0.4 and \SI{0.9}{\bohr}.
However, one should note that the small moment reported from \textsc{SQUID} measurements appears to be based on a high field saturation moment, which is similar in magnitude to the typical magnetization observed at 70~KOe (see Fig.~\ref{fig:Sub_All}).
Billington \emph{et al.}~attributed these discrepancies to the spin-glass behavior observed in the Cantor alloy and/or an unconventional magnetic ground state.
Concerning the present study, the fact that effective moments extracted from modified Curie-Weiss fits are large compared to the magnetization at 70~kOe (and large compared to DFT calculated moments) is fully consistent with the picture that the Cantor alloy is a Stoner-enhanced itinerant magnet~\cite{elmslie_magnetic_2022}.
Such systems can exhibit very large ratios of the effective moment to the saturation moment~\cite{rhodes_effective_1963}.

Finally, the moments extracted from our linear regression analysis and tabulated in the first row of Table~\ref{tab:peffs} should not be expected to correspond exactly to element specific moments directly determined via methods such as inelastic x-ray scattering.
Rather, they give the overall incremental contribution to the net effective moment as the concentration of one component of the alloy is adjusted.
As such, these moments provide a useful guide to efforts to design Cantor alloy variants with specific magnetic properties.

\section{Conclusions}
Near the equiatomic composition, the Cantor alloy undergoes two transitions, a ferrimagnetic transition at lower temperature and a spin-glass-like transition at higher temperature.
While the spin-glass-like transition temperature varies depending on concentration, the ferrimagnetic transition is remarkably stable, occurring at approximately \SI{43}{\kelvin} in all samples in which it appears despite not resulting from an impurity phase~\cite{elmslie_magnetic_2022}.
Trends in the magnetic properties are clearest in the \ch{Mn} samples, in which larger concentrations shift the spin-glass-like transitions to higher temperatures and decrease the net effective moment.
In alloys with low concentrations of \ch{Cr} or \ch{Mn}, however, the ferrimagnetic transition becomes ferromagnetic.

Fitting the susceptibility of Cantor alloy samples of many different compositions permits the estimation of the magnetic contribution from each constituent element by assuming a linear contribution from each.
This analysis reveals that larger concentrations of ferromagnetic elements \ch{Fe}, \ch{Co}, and \ch{Ni} increase the net effective magnetic moment of the Cantor alloy, while increased quantities of antiferromagnetic elements \ch{Cr} and \ch{Mn} decrease the net effective moment.
The effective magnetic moments extracted via this process provide a road map facilitating a fine level of control over the magnetic moment of the Cantor alloy by varying its composition.
Furthermore, similar regression analyses of data for a variety of chemically substituted samples could be extended to a wider range of high-entropy materials to provide similar insight into the composition dependence of a other properties.

\section{Declaration of Competing Interest}
The authors declare that they have no known competing financial interests or personal relationships that could have appeared to influence the work reported in this paper.

\section{Acknowledgments} \label{ack}
T.A.E.~and M.W.M.~acknowledge support from NSF DMR-1708410.
Synthesis and characterization facilities at the University of Florida were developed under support from NSF-CAREER 1453752 (J.J.H.).  
B.A.F.~and E.Z.~acknowledge support from the College of Physical and Mathematical Sciences at Brigham Young University and also thank the scientific staff at the Centre for Molecular and Materials Science at TRIUMF, particularly  Dr.~Gerald Morris, for support during the muon spin relaxation experiment.
A portion of this work was performed at the National High Magnetic Field Laboratory, which is supported by the National Science Foundation Cooperative Agreement No.~DMR-1644779 and the State of Florida.
The authors acknowledge the staff of the Nano Research Facility at University of Florida for their assistance and guidance in acquiring SEM and XRD data.

The computational work was performed, in part, at the Center for Integrated Nanotechnologies, an Office of Science User Facility operated for the U.S.\ Department of Energy.
Sandia National Laboratories is a multimission laboratory managed and operated by National Technology \& Engineering Solutions of Sandia, LLC, a wholly owned subsidiary of Honeywell International Inc., for the U.S.\ Department of Energy’s National Nuclear Security Administration under contract DE-NA0003525.
This paper describes objective technical results and analysis. Any subjective views or opinions that might be expressed in the paper do not necessarily represent the views of the U.S.\ Department of Energy or the United States Government.

\bibliography{references}

\end{document}


\begin{frontmatter}
\title{Supplementary Material:\\ Tuning the Magnetic Properties of the CrMnFeCoNi Cantor Alloy}
\author[1]{Timothy A.\ Elmslie}
\author[2]{Jacob Startt}
\author[3]{Yang Yang}
\author[3]{Sujeily Soto-Medina}
\author[5]{Emma Zappala}
\author[1,4]{\\Mark~W.\ Meisel}
\author[3]{Michele V.\ Manuel}
\author[5]{Benjamin A. Frandsen}
\author[2]{R\'emi Dingreville}
\author[1]{James J.\ Hamlin}
\ead{jhamlin@ufl.edu}

\address[1]{Department of Physics, University of Florida, Gainesville, FL 32611-8440, USA}
\address[2]{Center for Integrated Nanotechnologies, Sandia National Laboratories, Albuquerque, NM, 87185, USA}
\address[3]{Department of Materials Science and Engineering, University of Florida, Gainesville, FL 32611-6400, USA}
\address[5]{Department of Physics and Astronomy, Brigham Young University, Provo, UT 84602, USA}
\address[4]{National High Magnetic Field Laboratory, University of Florida, Gainesville, Florida 32611-8440, USA}

\date{\today}
\journal{Acta Mater.}
\end{frontmatter}
\section{Compositional and Structural Analysis}
\begin{figure}[t]
    \centering
     \includegraphics[width=0.9\textwidth]{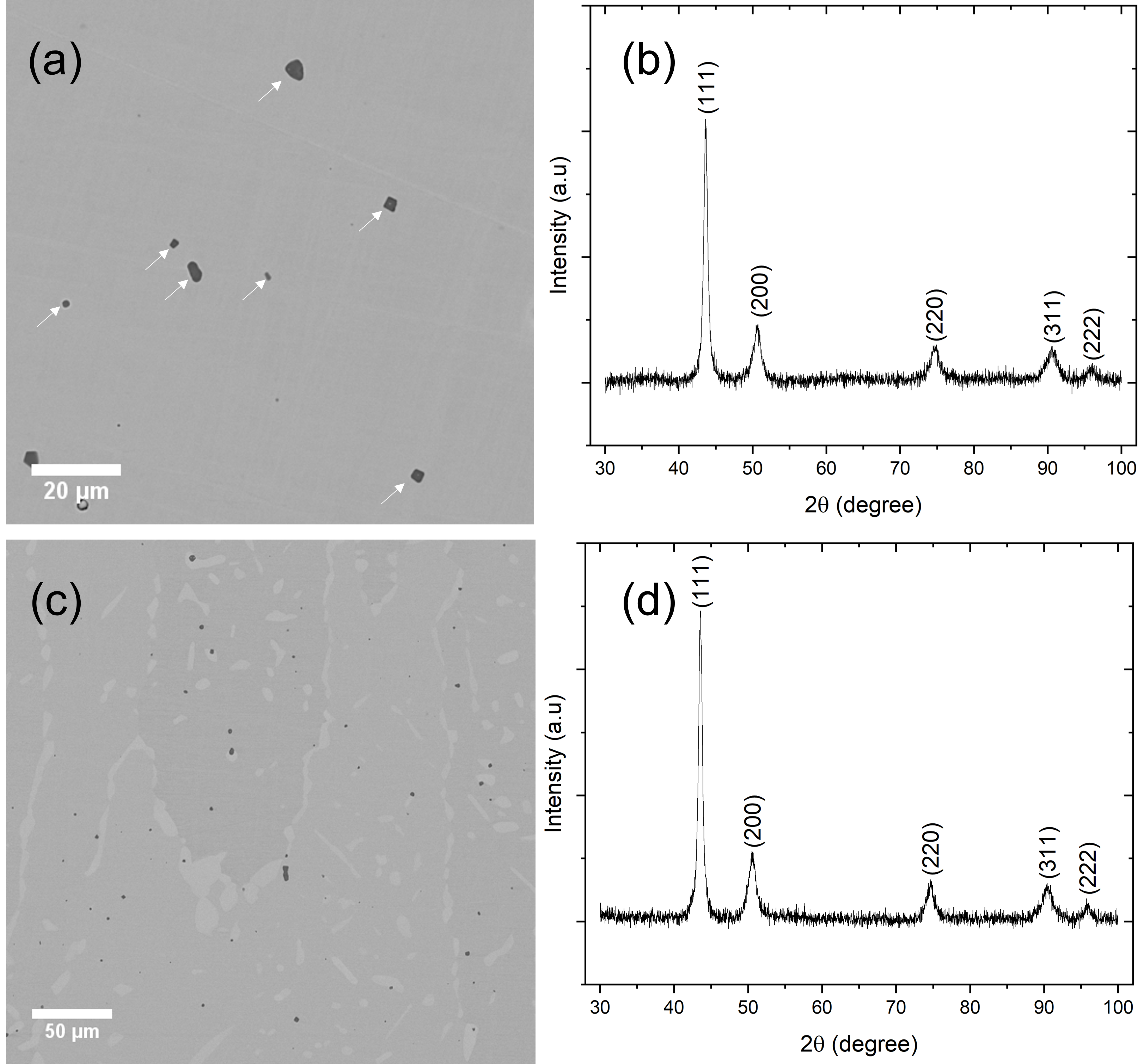}\label{fig:supplement Ni10&Cr30}
 
     \caption{(a) SEM-BSE image of the anneal Ni$_{10}$ contains small number of oxide particles indicated by arrows. The compositions of the inclusions were found rich in Cr and Mn using EDX. (b) The crystal structure of the Ni$_{10}$ was found FCC single phase using XRD.  (c) SEM-BSE image of the anneal Cr$_{30}$ showed oxide particles, pores and the second phase with lighter contrast. (d) The second phase was not detected using XRD, which showed FCC single phase.}
\end{figure}

The representative samples Ni$_{10}$ and Cr$_{30}$ were investigated using SEM and XRD. In the Ni$_{10}$ sample, micro-sized pores were found, which are considered formed during casting. 
Using BSE and EDX, a small number of particles were detected and identified as oxides rich in Cr and Mn. 
The crystal of the sample was determined as FCC single phase according to the XRD shown in Fig.~\ref{fig:supplement Ni10&Cr30}. No peak corresponding to the oxide particles was found in the XRD pattern. 
This kind of (Cr, Mn)-rich particle was reported in previous studies~\cite{otto_microstructural_2014,pickering_precipitation_2016,otto_decomposition_2016}, but was not able to be detected using XRD. 
The investigated properties were unlikely to be affected by these inclusions due to the small volume fraction thereof. 
The compositions of the anneal Ni$_{10}$ was measured using EDX  as shown in Table~\ref{tab:Composition Ni10}.
\begin{table}[htbp]
  \caption{Compositions of the anneal \ch{Ni_{10}} measured by SEM-EDX.}
  \begin{center}
  \begin{tabular}{c c}
  \hline
  Element & at.\%\\
  \hline
  \ch{Cr} & 22(1) \\
  \ch{Mn} & 22.1(2) \\
  \ch{Fe} & 22.9(1) \\
  \ch{Co} & 22.9(9) \\
  \ch{Ni} & 9.7(3) \\
  \hline
 \end{tabular}
  \end{center}
 \label{tab:Composition Ni10}
 \end{table}

Figure~\ref{fig:supplement Ni10&Cr30} shows SEM-BSE and XRD measurements on the samples Ni$_{10}$ and Cr$_{30}$.
In the Cr$_{30}$ sample, a secondary phase was observed in the BSE image with lighter contrast. This secondary phase was inhomogeneously distributed along the grain boundaries, pores and inclusions. The composition of this phase is characterized rich in Cr than the matrix using EDX as shown in Table~\ref{tab:Composition Cr30}. However, no peak was found in the XRD pattern corresponding to the second phase.

\begin{table}[htbp]
  \caption{Compositions of the second phase and matrix in anneal \ch{Cr_{30}} measured by SEM-EDX.}
  \begin{center}
  \begin{tabular}{c c c}
  \hline
   Element at.\%\ & Precipitate & Matrix \\
  \hline
  \ch{Cr} & 46.5(1) & 29.5(1)\\
  \ch{Mn} & 14.3(1) & 17.9(1)\\
  \ch{Fe} & 15.5(1) & 18(1)\\
  \ch{Co} & 14.7(1) & 17.4(1)\\
  \ch{Ni} & 8.9(1) & 17.2(1)\\
  \hline
 \end{tabular}
  \end{center}
 \label{tab:Composition Cr30}
 \end{table}

\clearpage
\section{Modified Curie-Weiss Fits}
\begin{figure}[h]
    \centering
    \includegraphics[height=0.7\textheight]{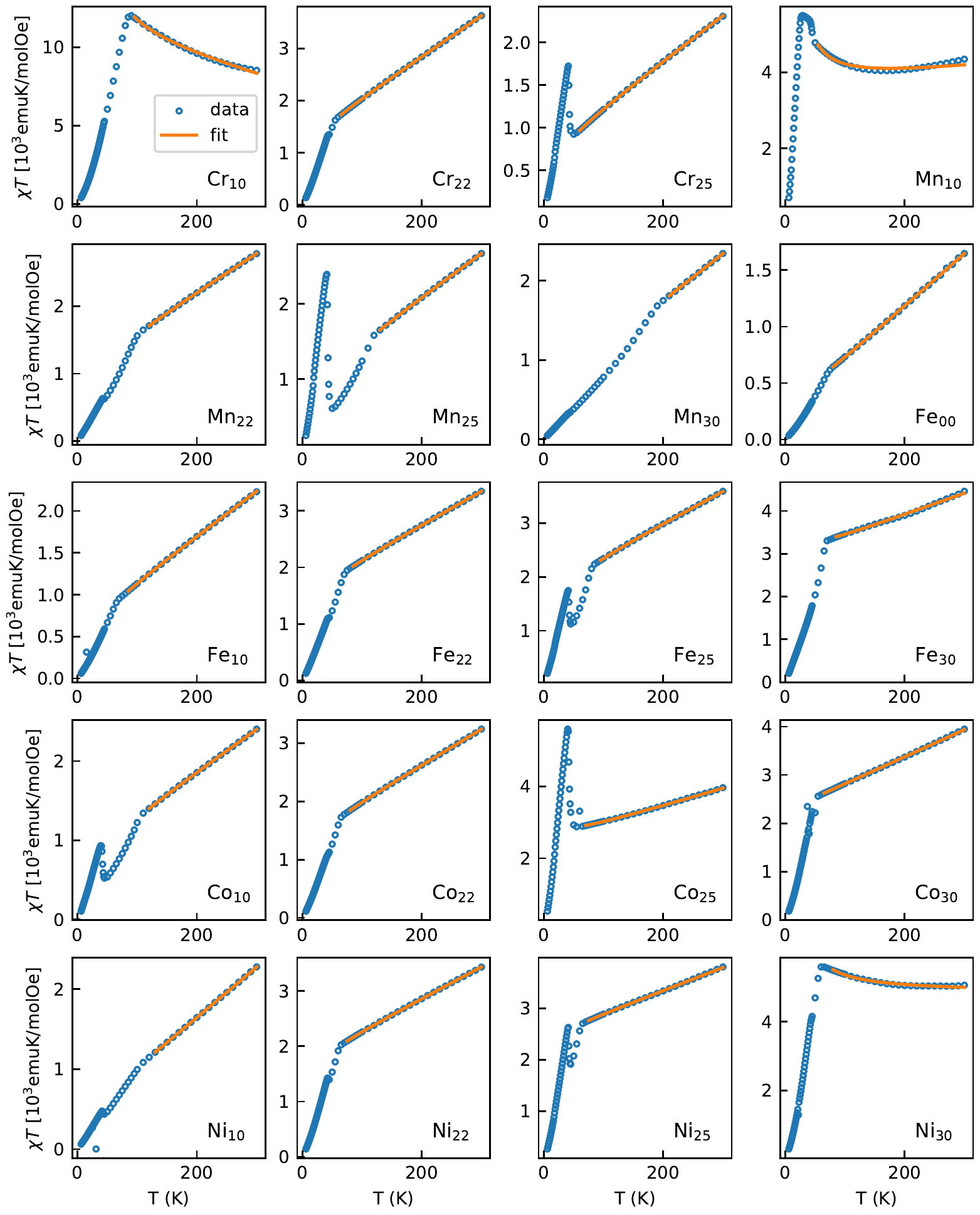}
    \caption{Modified Curie-Weiss fits shown in orange compared with experimental data shown in blue for a characteristic selection of measured samples.  Data are plotted as susceptibility times temperature versus temperature for greater readability across the entire measured temperature range.}
    \label{fig:allFits}
\end{figure}

The $\chi_0$, $p_{\rm{eff}}$, and $\Theta$ values displayed in Table~1 (main text) were obtained by fitting a modified Curie-Weiss equation to experimental susceptibility data.
We used the \texttt{scipy.optimize.curve\_fit} non-linear least squares fitting function from the scipy library to perform the fits.
The datasets and corresponding fits for each fitted sample are compared in Fig.~\ref{fig:allFits}.

\clearpage
\section{Least-squares solution for element-specific magnetic moments}
As described in the main text we construct a system of equations (Eq.~3 in the main text) using the measured effective moments and known elemental concentrations for each sample.
The unknowns in these equations are the element-specific magnetic moments.
It is possible to write this equation for each sample.
We then solved the linear matrix equation using the \texttt{numpy.linalg.lstsq} function from the numpy library.
Since the system of equations is over-determined, we obtain best fit values for the element-specific moments, rather than an exact solution.

\subsection{Omitted samples}
Three samples that are included in Table~1 (main text) were omitted from this analysis.
These included the \ch{Cr0}, \ch{Cr10}, and \ch{Mn0} samples.
After omitting these samples, data for 23 samples remained.
This included four distinct samples of the equiatomic composition.
In the case of \ch{Cr0} and \ch{Mn0}, fitting with the modified Curie-Weiss equation (Eq.~1 main text), produced unreliable results that depended on the temperature range of the fit or produced unphysical large negative values of $\chi _0$.
This is presumably related to transition to a more strongly ferromagnetic state in those materials and, in the case of \ch{Cr0} the fact that a very limited temperature range above the Curie temperature is available for fitting.

The \ch{Cr10} magnetization data was relatively well fit by the modified Curie-Weiss equation as shown in the top row of Fig.~\ref{fig:allFits}.
However, the fit produced an anomalously large $p_{\rm{eff}} \sim 3 \mathrm{\mu _B}$.
Including the \ch{Cr10} data in the global least squares solution produced overall poor agreement.
We found that leaving out the \ch{Cr10} data reduced the overall residual of the fit by nearly a factor of four, while leaving out any other data for any other sample had a minimal effect on the residual.
We therefore omitted the \ch{Cr10} datapoint from the fit.
The reason for the poor agreement of the \ch{Cr10} point with the rest of the data is likely connected with the fact that this sample is on the verge of a transition to the more ferromagnetic state that is observed in \ch{Cr0}
We are currently further investigating the transition to ferromagnetism in the Cr-poor series.

\subsection{Uncertainty estimates}
\begin{figure}
    \centering
    \includegraphics[width=\textwidth]{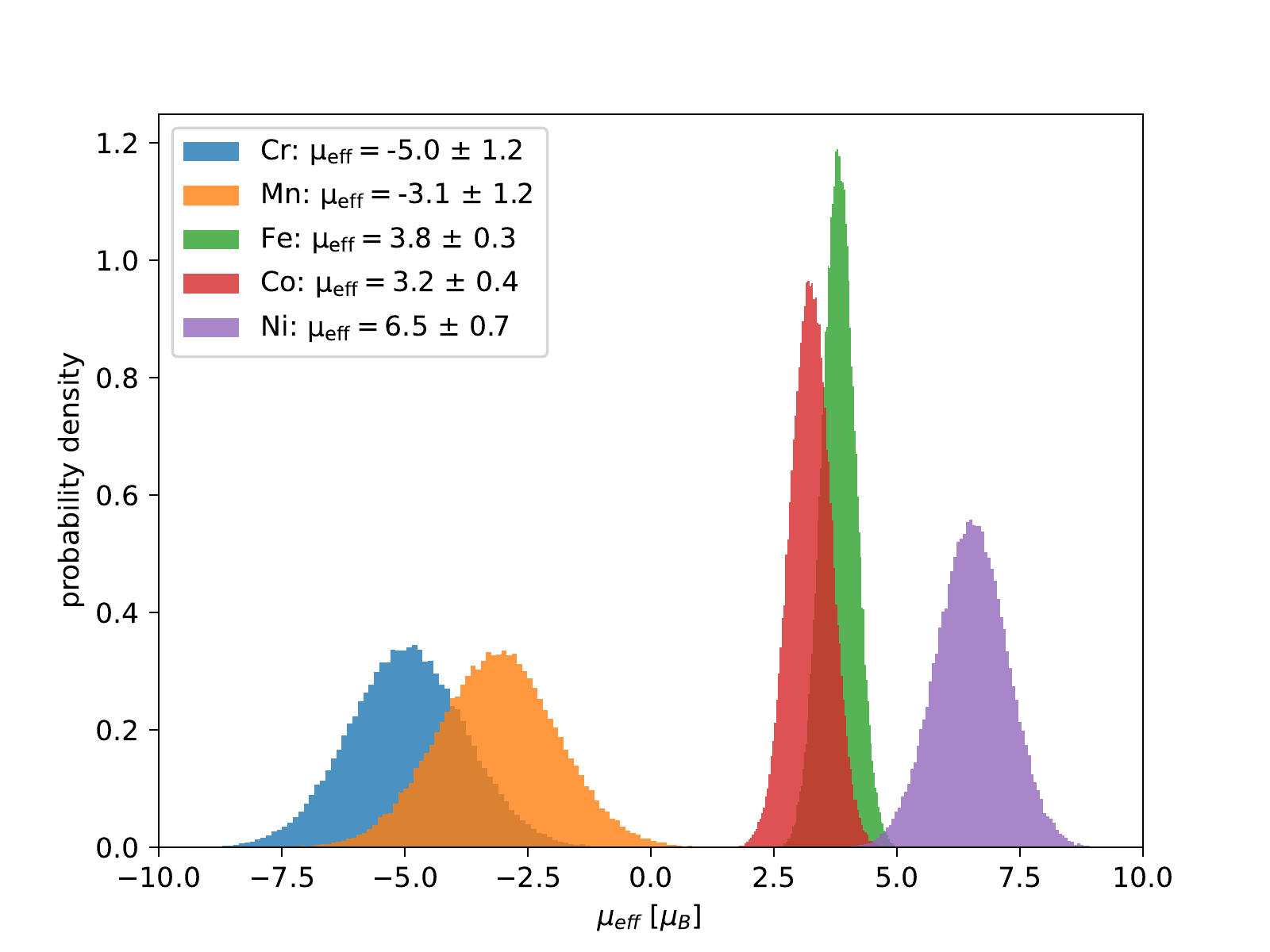}
    \caption{Distribution of element-specific magnetic moments as determined by randomly varying the values of the input $p_{\rm{eff}}$ values.  This allows the uncertainty in the values of $p_{\rm{eff}}$ for each sample to be accounted for in determining the uncertainties in the element-specific magnetic moments.}
    \label{fig:uncertainty}
\end{figure}

The \texttt{numpy.linalg.lstsq} function does not provide a straightforward means of taking into account the uncertainties in the values of $p_{\rm{eff}}$ for each sample (left side of Eq.~3 in the main text).
In order to generate uncertainty values for the element-specific moments that take into account the uncertainties in the input ($p_{\rm{eff}}$) data, we used the following approach:
\begin{enumerate}
    \item Apply a Gaussian randomization to the values of $p_{\rm{eff}}$ such that they vary around the best fit value with a standard deviation equal to the uncertainties reported in Table~1 (main text).
    \item Solve the resulting linear matrix equation consisting of data on 23 samples to produce a set of element-specific magnetic moments. 
    \item Repeat steps 1-2 a total of $10^5$ times to produce a distribution of element-specific moments.
    These distributions are shown in Fig.~\ref{fig:uncertainty}
    \item The values and uncertainties of the element-specific magnetic moments are then taken from the mean and standard deviations of the resulting distributions.
\end{enumerate}

\bibliography{references}